\documentclass{article}
\usepackage{spconf}

\usepackage{bbm}
\usepackage{amssymb,amsmath}
\usepackage{xcolor}
\usepackage{booktabs} 
\usepackage{algorithm}
\usepackage[noend]{algpseudocode}

\usepackage{url}
\usepackage{graphicx}
\usepackage{caption}
\usepackage{subcaption}

\newcommand\an[1]{\textcolor{magenta}{}}
\newcommand\as[1]{\textcolor{teal}{}}
\newcommand\yk[1]{\textcolor{red}{}}
\newcommand\gh[1]{\textcolor{violet}{}}
\newcommand\sm[1]{\textcolor{brown}{}}
\newcommand\ourmethod{AdaKWS}

\title{Open-vocabulary Keyword-spotting with Adaptive Instance Normalization}
\name{Aviv Navon, Aviv Shamsian, Neta Glazer, Gill Hetz, Joseph Keshet} 
\address{aiOla Research}
\begin{document}
%
\maketitle
\begin{abstract}

Open vocabulary keyword spotting is a crucial and challenging task in automatic speech recognition (ASR) that focuses on detecting user-defined keywords within a spoken utterance. Keyword spotting methods commonly map the audio utterance and keyword into a joint embedding space to obtain some affinity score. 
In this work, we propose \textit{\ourmethod{}}, a novel method for keyword spotting in which a text encoder is trained to output keyword-conditioned normalization parameters. These parameters are used to process the auditory input. We provide an extensive evaluation using challenging and diverse multi-lingual benchmarks and show significant improvements over recent keyword spotting and ASR baselines. Furthermore, we study the effectiveness of our approach on low-resource languages that were unseen during the training. The results demonstrate a substantial performance improvement compared to baseline methods.

\end{abstract}
\section{Introduction}
\label{sec:intro}


Keyword spotting (KWS) is a critical component of many speech recognition systems~\cite{lopez2021deep}. It involves identifying specific words or phrases within a continuous audio stream. KWS is essential for various applications, from automated transcription to voice-activated assistants~\cite{fuchs2021cnn}. 
Despite recent advancements, KWS still faces unresolved challenges in adaptability and customization.
Keyword spotting methods are frequently trained to detect a predefined set of target keywords~\cite{berg2021keyword,majumdar2020matchboxnet,ding2022letr}. This requires a large amount of labeled data per keyword and widely restricts the usability of the models. Importantly, adapting to custom target keywords generally requires retraining or finetuning. Recently, few-shot learning and query-by-example approaches have arisen as a more flexible alternative for KWS~\cite{reuter2023multilingual,lee2023fully,kirandevraj2022generalized,jung2023metric, Lugosch2018DONUTCQ, Huang2021QueryByExampleKS, sacchi2019open}. However, these works are still limited in performance on novel out-of-vocabulary keywords and require obtaining several auditory examples, which can be restricting and challenging, specifically in low-resource languages or domains.

A fully adaptable keyword-spotting approach that can accurately detect user-defined keywords in multiple languages, with no additional examples or optimization, remains a challenging goal in KWS. Recently, several methods have proposed open-vocabulary KWS systems that can generalize to keywords not seen during training~\cite{shin2022learning,nishu2023matching,nishu2023flexible}.
The methods rely on a text encoder for aligning audio and text information in a joint latent space. While these methods eliminate the need for auditory examples for novel keywords, they face several limitations. First, a misalignment in audio and text representations may arise from using two disjoint encoders~\cite{nishu2023flexible}. \an{Yossi, please approve} Second, the reliance on a phoneme model can limit the applicability in low-resource languages.
Lastly, these methods are evaluated on English 
benchmarks, and it is unclear how these approaches generalize to diverse languages and dialects.

\begin{figure*}[t!]
\begin{subfigure}[b]{0.55\linewidth}
\centering
\includegraphics[width=\textwidth]{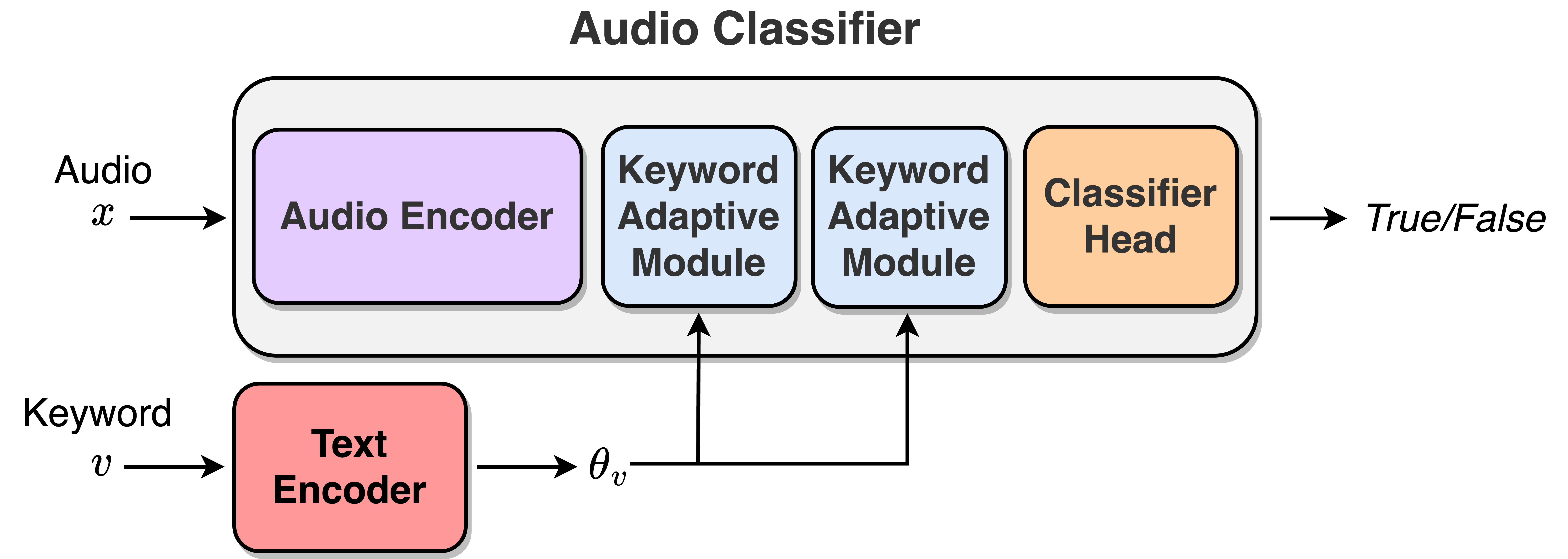}
\caption{}
\label{fig:archi_main}
\end{subfigure}
\begin{subfigure}[b]{0.45\linewidth}
\centering
\includegraphics[width=\textwidth]{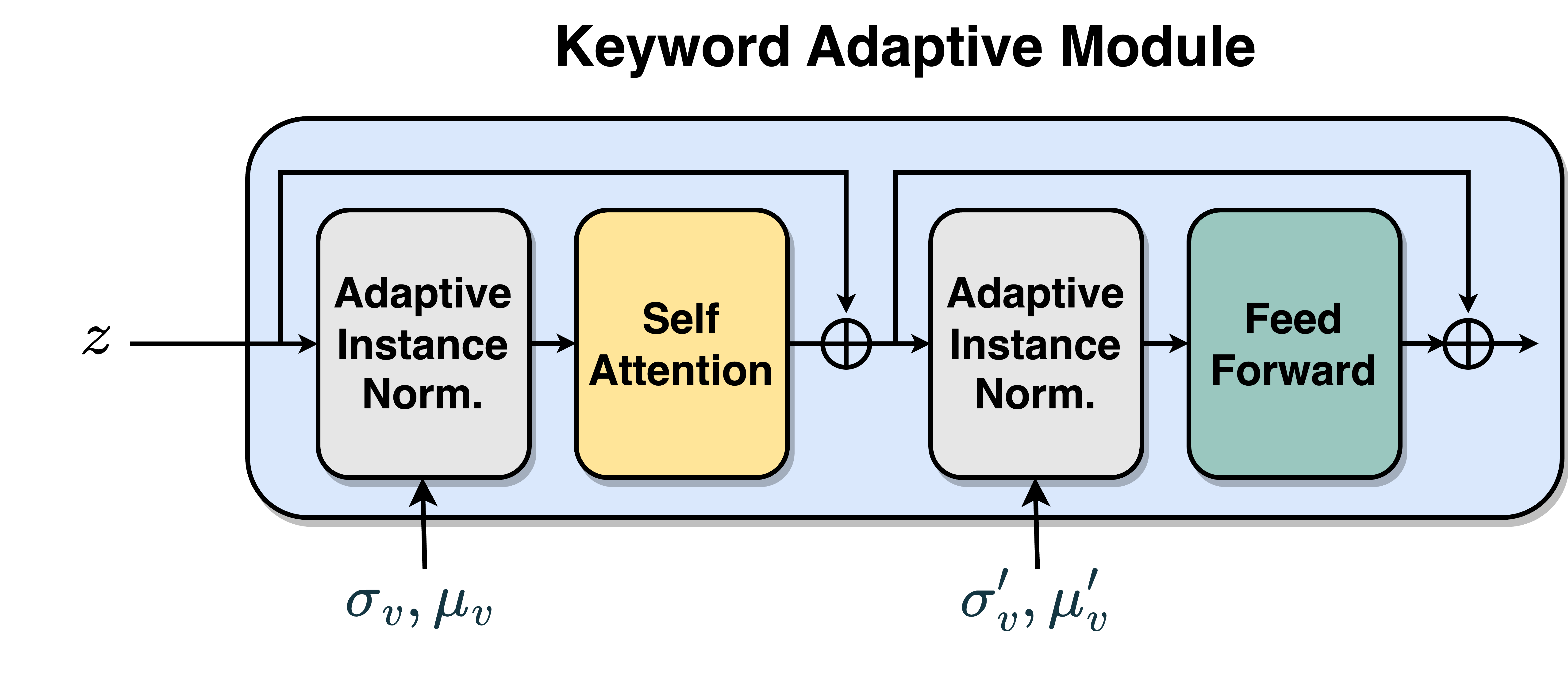}
\caption{}
\label{fig:archi_ada}
\end{subfigure}
\caption{The architecture for \ourmethod{}: (a) 
Overview: A text encoder outputs a set of keyword-conditioned normalization parameters. 
The audio is processed using the classifier which is conditioned on the keyword through the keyword-adaptive modules. (b) A detailed view of the keyword adaptive module.}
\label{fig: kws}
\end{figure*}

To address these issues, we take a different path to the open-vocabulary problem and propose a novel approach for KWS, termed \ourmethod{}. Instead of embedding textual and auditory information into a joint latent space, we employ a character-based LSTM encoder which maps an input keyword into a set of normalization parameters. These parameters are used to process the audio signal in keyword-adaptive modules (see, Figure~\ref{fig: kws}). The adaptive module is a standard transformer encoder module in which we swap the Layer Normalization layers with an Adaptive Instance Normalization (AdaIN) layers~\cite{Huang2017ArbitraryST}, where the adaptive parameters are keyword-conditioned. AdaIN layers have proven highly effective for multiple tasks and domains, including image harmonization, image-to-image translation, and style editing. 


To reduce the false detection of acoustically similar keywords and effectively train our model, we introduce a new technique for mining \textit{hard} negative examples. Importantly, unlike previous recent works~\cite{shin2022learning,nishu2023matching,nishu2023flexible} which train the KWS model on segmented audio samples containing the keyword alone, we train our model on entire sentences (up to 30 seconds). This eliminates the need for word-level alignment or expensive preprocessing and dramatically increases the amount of available training data.

To summarize, we make the following contributions: (1) Propose a novel KWS approach, named \ourmethod{}, based on adaptive instance normalization layers, (2) Introduce a new technique for mining hard negative keyword examples, (3) Demonstrate the benefits and effectiveness of our approach using a diverse set of multilingual benchmarks, and generalization to unseen languages.



\section{Keyword-spotting using Adaptive Instance normalization}
\label{sec:method}


The \ourmethod{} model is constructed of two main building blocks. A text encoder, and an audio classifier. The audio classifier is in turn construct of two main modules, an audio encoder, and a keyword-adaptive module
(see Figure \ref{fig: kws}).

We first focus on the audio classifier, $f$. As the audio encoder, we use a pre-trained Whisper transformer encoder~\cite{Radford2022RobustSR}. In our experiments, we keep the audio encoder frozen. The audio is first processed using the audio encoder, and the resulting audio representation is then passed to two sequential keyword-adaptive modules. Each module is a standard transformer encoder block in which we swap the Layer Normalization layers with Adaptive Instance Normalization layers (AdaIN). An AdaIn is a normalization layer of the form, 
\begin{equation}\label{eq:adain}
\text{AdaIN}(z,v)=\sigma_v\left(\frac{z-\mu_z}{\sigma_z}\right)+\mu_v.
\end{equation}
In our case, $z$ is the audio representation and $v$ is the target keyword. Finally, the keyword-conditioned audio representation is max-pooled and fed into a linear classifier. 
To map the target keyword to the corresponding normalization parameters, we employ a lightweight text encoder, $h$. The text encoder is a 4-layers character-based LSTM with a $256$ hidden dimension.

More formally, let $\phi$ denote the parameters of the text encoder $h$, and let $\theta$ denote the collection of parameters in $f$ which are shared among keywords (i.e., all parameters that are not in the AdaIN layers). To detect a keyword $v$ in an auditory utterance $x$, we first process $v$ using $h$ to obtain the keyword conditioned normalization parameters $h(v;\phi)=\theta_v$. 
Here, $\theta_v$ denote collection of all normalization parameters $\mu(v), \sigma(v)$ from Eq.~\ref{eq:adain}
Then, we determine the existence of $v$ in $x$ according to $f(x;\theta,\theta_v):=P(v|x)$.

Note that under the above setup, the trainable parameters are $\phi, \theta$. The text encoder parameters can be updated according to the chain rule,
\begin{equation}
\nabla_{\phi}\mathcal{L}(x,v)=(\nabla_\phi\theta_v)^T\nabla_{\theta_v}\mathcal{L}(x,v)
\end{equation}
where $\mathcal{L}$ denotes the classification loss (i.e., cross-entropy) for $(x,v)$. We summarize our training procedure in Alg.~\ref{alg:kws-hn}.


Our method, \ourmethod{}, provides a natural way for sharing information across keywords, through the shared parameters $(\phi, \theta)$, while maintaining the flexibility of generating diverse keyword-specific classifiers.

\begin{algorithm}[t]
    \centering
    \caption{\ourmethod{}}\small\label{alg:kws-hn}
    \begin{algorithmic}[t]
    \State \textbf{input:} $\alpha$ --- Text encoder learning rate, $\eta$ --- classifier learning rate, $\theta$ --- Audio classifier parameters, $\phi$ --- Text encoder parameters
    \For{$r=1,...,n$}
    \State sample a batch of pos/neg examples $\{(x_i, v_i)\}_{i=1}^B.$
    \State compute $\theta_{v_i}=h(v_i;\phi)$
    \State $\theta=\theta-\eta \frac{1}{B}\sum_i \nabla_{\theta}\mathcal{L}(x_i,v_i)$
    \State $\phi=\phi-\alpha \frac{1}{B}\sum_i (\nabla_{\phi}\theta_{v_i})^T \nabla_{\theta_{v_i}}\mathcal{L}(x_i,v_i)$
    \EndFor
    \State \textbf{return:} $\phi, \theta$
    \end{algorithmic}
\end{algorithm}

\begin{table*}[!t]
\centering
    \caption{\textit{VoxPopuli:} F1 results for the VoxPopuli test dataset.}
    \vskip 0.11in
\tiny
\begin{tabular}{lcccccccccccccccccc}
\toprule
 &  CS & DE & EN & ES & ET & FI & FR & HR & HU & IT & LT & NL & PL & RO & SK & SL & Overall &  \# Params \\
\midrule
Whisper-Tiny & $48.5$ & $77.0$ & $83.4$ & $85.6$ & $39.5$ & $58.4$ & $78.3$ & $48.6$ & $47.0$ & $72.2$ & $38.4$ & $67.8$ & $74.7$ & $52.0$ & $43.8$ & $46.0$ & $69.3$ &$39$M \\
Whisper-Small & $77.9$ & $89.3$ & $ 84.1 $ & $86.3$ & $55.2$ & $81.5$ & $90.5$ & $72.5$ & $72.5$ & $85.4$ & $62.5$ & $85.8$ & $89.7$ & $77.3$ & $68.8$ & $64.3$ & $83.3$ & $244$M \\
Whisper-Large-V2 & $91.0$ & $93.2$ & $80.5$& $87.8$ & $78.5$ & $89.1$ & $93.2$ & $67.9$ & $87.4$ & $87.7$ & $76.1$ & $\mathbf{92.6}$ & $93.8$ & $87.7$ & $84.0$ & $80.0$ & $88.4$ & $1550$M  \\
\midrule
\ourmethod{}-Tiny &  $91.8$ & $94.4$ & $95.5$ & $95.1$ & $86.2$ & $92.1$ & $94.9$ & $89.3$ & $90.6$ & $90.8$ & $\mathbf{85.6}$ & $91.1$ & $93.0$ & $91.9$ & $91.7$ & $85.6$ & $92.8$ & $15$M  \\
\ourmethod{}-Base & $92.3$ & $95.2$ & $95.9$ & $95.1$ & $85.2$ & $92.2$ & $\mathbf{96.3}$ & $90.4$ & $91.3$ & $92.0$ & $82.1$ & $91.5$ & $94.3$ & $92.9$ & $93.9$ & $88.5$ & $93.7$ & $31$M \\
\ourmethod{}-Small & $\mathbf{94.4}$ & $\mathbf{95.7}$ & $\mathbf{96.3}$ & $\mathbf{95.6}$ & $\mathbf{91.1}$ & $\mathbf{94.8}$ & $95.9$ & $\mathbf{93.0}$ & $\mathbf{92.7}$ & $\mathbf{92.8}$ & $78.5$ & $92.2$ & $\mathbf{95.5}$ & $\mathbf{94.8}$ & $\mathbf{94.5}$ & $\mathbf{89.9}$ & $\mathbf{94.6}$ & $109$M \\
\bottomrule
\end{tabular}
\label{tab:voxpopuli}
\end{table*}

\section{Negative Sampling}
\label{sec:negsampling}

In this section, we present our negative sampling approach which we employ for creating a diverse set of \textit{hard} negative examples per batch.

Let $\{(x_i,\mathbf{v}_i)\}$ for $i=1,...,N$ denotes the training data. Here $x$ denotes the speech utterance and $\mathbf{v}=(v_1,...,v_M)$ denotes the corresponding transcripts, with $v_i\in\mathcal{V}$ for all $i$. Here $\mathcal{V}$ denotes a set of training keywords. During training, given an example $(x,\mathbf{v})$, we sample a random positive keyword $v^+\in \mathbf{v}$ to form a positive training example $(x, v^+)$. In order to generate a negative training example, we consider several alternatives.\\
\textbf{Random negative.} Here we simply sample a random keyword from $v^-\in \mathcal{V}\setminus \mathbf{v}$. We empirically found that this simple approach is not sufficient for training a KWS model that can accurately separate acoustic similar words. Intuitively, the random sampled keywords are acoustically far from the keywords in $\mathbf{v}$.

To address this issue, we introduce several approaches for constructing hard negative examples.\\
\textbf{Character substitution.} Here, we alter a positive keyword $v^+$, by substituting one or more of its characters. The new character can be chosen randomly, or according to an a priori mapping of acoustically similar characters (``s"$\to $``z",``p"$\to$``b", etc.).\\
\textbf{Keyword concatenation.} Here we form a negative keyword by concatenating a random keyword $v\in \mathcal{V}\setminus \mathbf{v}$ to a positive keyword $v^+$, i.e., $v^-=v\circ v^+$ or $v^-=v^+\circ v$, where $\circ$ denotes the concatenation operator.\\
\textbf{Nearest keyword (NK).} To obtain negative keyword $v^-$ that are acoustically similar to a reference keyword $v^+$, we sample $v^-$ according to the text embedding representation $e(v)$. Specifically, we use the last hidden layer of $h$ to form the embedding representation. To ensure efficiency, we sample negative examples within each training batch, by querying for the keyword with the smallest cosine distance, i.e., $v^-=\arg\min_{v\in \mathcal{V}_\mathcal{B}\setminus v^+} d(e(v^+),e(v))$, where $\mathcal{V}_\mathcal{B}$ denotes the keywords in batch $\mathcal{B}$.

\begin{figure}
    \includegraphics[width=1.\columnwidth]{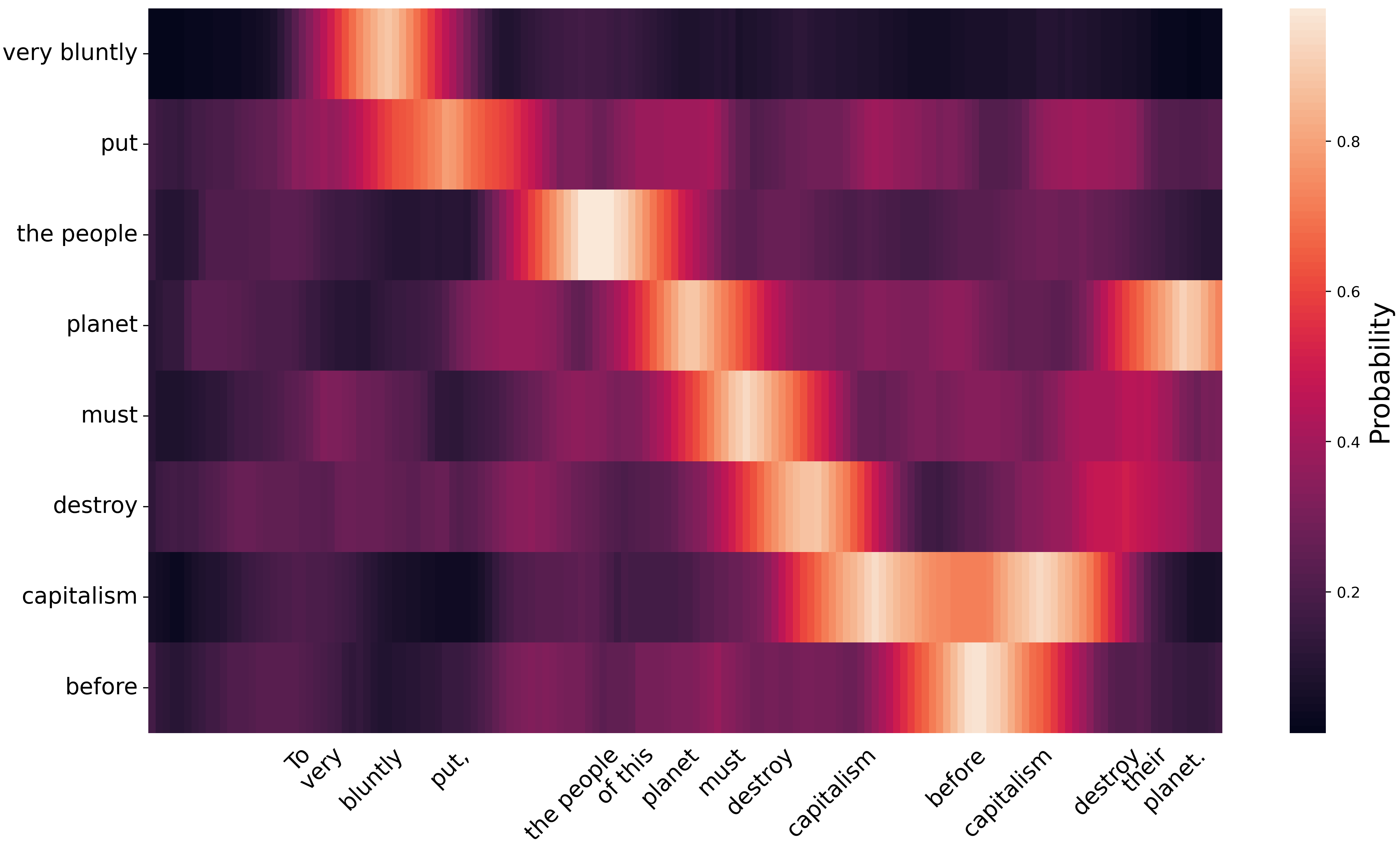}
    \caption{\textit{Model predictions}. We illustrate the predicted probability of a queried keyword (y-axis) across different time windows.  Note that higher probability scores are aligned with the temporal occurrence of the word within the utterance (x-axis).}
    \label{fig:pred_align}
\end{figure}


\begin{table}[ht]
\small
\centering
    \caption{\textit{LibriPhrase}: AUC and EER results for LibriPhrase hard (LH) and easy (LE) test splits.}
    \vskip 0.11in
\begin{tabular}{lcccc}
\toprule
 &  \multicolumn{2}{c}{AUC (\%) $\uparrow$} & \multicolumn{2}{c}{EER (\%) $\downarrow$}\\
 \cmidrule(lr){2-3} \cmidrule(lr){4-5} 
 & LH & LE & LH & LE \\
\midrule
Triplet~\cite{sacchi2019open} & $54.88$ & $63.53$ & $44.36$ & $32.75$\\
Attention~\cite{Huang2021QueryByExampleKS} & $62.65$ & $78.74$ & $41.95$ & $28.74$\\
DONUT~\cite{Lugosch2018DONUTCQ} & $54.88$ & $63.53$ & $44.36$ & $32.75$\\
CMCD~\cite{shin2022learning} & $73.58$ & $96.70$ & $32.90$ & $8.42$ \\
EMKWS~\cite{nishu2023matching} & $84.21$ & $97.83$ & $23.36$ & $7.36$ \\
CED~\cite{nishu2023flexible} & $92.70$ & $\mathbf{99.84}$ & $14.40$ & $1.70$ \\
\midrule
\ourmethod{}-Tiny & $93.75$ & $99.80$ & $13.47$ &  $1.61$ \\
\ourmethod{}-Base & $94.39$ & $99.81$ & $12.60$ &  $1.37$ \\
\ourmethod{}-Small & $\mathbf{95.09}$ & $99.82$ & $\mathbf{11.48}$ & $\mathbf{1.21}$ \\
\bottomrule
\end{tabular}
\label{tab:libriphrase}
\end{table}

\section{Experiments}
\label{sec:exp}

\begin{table*}[!t]
\small
\centering
    \caption{\textit{Multilingual-LibriSpeech:} Zero-shot F1 results for the \textit{Multilingual-LibriSpeech} test dataset.}
    \vskip 0.11in
\begin{tabular}{lcccccccccc}
\toprule
 &  DE & EN & ES & FR & IT & NL & PL & Overall &  \# Params & Inference Time (MS) $\downarrow$ \\
\midrule
Whisper-Tiny & $77.6$ & $85.8$ & $83.6$ & $73.4$ & $72.4$ & $69.2$ & $76.0$ &  $77.9$ & $39$M & $260$ \\
Whisper-Small & $90.8$ & $91.3$ & $92.9$ & $88.2$ & $85.8$ & $86.1$ & $91.5$ & $89.6$ & $244$M & $621$ \\
Whisper-Large-V2 & $\mathbf{95.0}$ & $\mathbf{93.8}$ & $\mathbf{95.1}$ & $93.8$ & $91.7$ & $\mathbf{92.3}$ & $\mathbf{95.2}$ & $\mathbf{93.9}$ & $1550$M  & $1836$ \\
\midrule
\ourmethod{}-Tiny & $92.6$ & $91.4$ & $92.1$ & $91.9$ & $92.3$ & $87.9$ & $92.6$ &  $91.3$ & $15$M & $6$ \\
\ourmethod{}-Base & $93.8$ & $92.6$ & $92.2$ & $92.8$ & $92.7$ & $90.0$ & $92.8$ & $92.4$ & $31$M & $7$\\
\ourmethod{}-Small & $94.6$ & $\mathbf{93.8}$ & $94.4$& $\mathbf{94.1}$ & $\mathbf{94.2}$ & $91.8$ & $94.4$ & $93.8$ & $109$M & $11$ \\
\bottomrule
\end{tabular}
\label{tab:mls}
\end{table*}

In this section, we compare \ourmethod{} with various KWS approaches on different learning setups and datasets. The experiments show the superiority of \ourmethod{} over previous KWS methods. 
\\
\textbf{Datasets.} We use several datasets.
\textit{VoxPopuli}~\cite{Wang2021VoxPopuliAL}: A large-scale multilingual speech corpus collected from the European parliament, with $\sim 1800$ hours of transcribed utterances from $16$ languages. \textit{LibriPhrase}~\cite{shin2022learning}: A recent benchmark for KWS based on the LibriSpeech~\cite{Panayotov2015LibrispeechAA} dataset. It consists of two splits, LibriPhrase Hard (LH) and LibriPhrase Easy (LE). Furthermore, we evaluate our model using two additional datasets, \textit{Multilingual LibriSpeech} and \textit{Fluers}~\cite{Conneau2022FLEURSFL}. To construct our evaluation set, we first randomly sample positive keywords. Then, we sample random, concat, and swap negatives as described in Section~\ref{sec:negsampling}, with equal probability. This results in a diverse and challenging open-vocabulary KWS benchmark.\\
\textbf{Experimental setup.} The \ourmethod{} models are trained using the VoxPopuli dataset for $25$ epochs with batch-size $144$ and $1e-4$ learning rate. We train three variants with varying audio encoder sizes, Tiny, Base, and Small, corresponding to the frozen Whisper encoder (audio encoder, Figure~\ref{fig: kws}). We report common metrics in the KWS field, namely F1 score, Area Under the Curve (AUC), and Equal Error Rate (ERR). \\
\textbf{Data preprocessing.} We follow the data processing procedure from~\cite{Radford2022RobustSR}. Each audio instance is resampled to a frequency of 16KHz, then an 80-channel log-magnitude Mel spectrogram representation is generated using 25-millisecond windows and 10-millisecond strides.\\
\textbf{Baselines.} We compare \ourmethod{} with recent keyword spotting and ASR works. The compared methods include: (1) \textbf{Whisper} \cite{Radford2022RobustSR} - We transcribe the input audio into text and subsequently conduct a search to ascertain the presence of the desired keyword within the output text. (2) \textbf{CED}~\cite{nishu2023flexible} - flexible KWS method using audio-compliant text encoder. (3) \textbf{EMKWS}~\cite{nishu2023matching} - contrastive learning approach using text-audio embedding matching. (4) \textbf{CMCD}~\cite{shin2022learning} - fusing the text and audio representation to shared latent space using cross-modal attention.
Additional query by example methods (5) \textbf{DONUT}~\cite{Lugosch2018DONUTCQ}, (6) \textbf{Attention}~\cite{Huang2021QueryByExampleKS}, and (7) \textbf{Triplet}~\cite{sacchi2019open}.

\begin{table}[H]
\scriptsize
\centering
    \caption{\textit{Zero-shot performance for low resource, novel languages:} F1 results for four novel and low resources languages from the \textit{Fleurs} test dataset.}
    \vskip 0.11in
\begin{tabular}{lccccc}
\toprule
 &  Icelandic & Maltese & Swahili & Uzbek & Overall \\ 
\midrule
Whisper-Tiny & $42.3$ & $37.6$ & $43.9$ & $34.2$ & $37.8$ \\
Whisper-Small & $52.0$ & $34.9$ & $55.6$ & $35.4$ & $40.4$ \\
Whisper-Large-V2 & $69.2$ & $38.6$ & $\mathbf{74.9}$ & $36.6$ & $48.1$ \\
\midrule
\ourmethod{}-Tiny & $69.2$ & $\mathbf{76.7}$ & $70.1$ & $\mathbf{67.5}$ &  $\mathbf{71.9}$ \\
\ourmethod{}-Base & $\mathbf{71.5}$ & $\mathbf{76.7}$ & $71.6$ & $65.7$ &  $71.6$ \\
\ourmethod{}-Small & $70.6$ & $76.0$ & $69.4$ & $66.1$ & $70.9$ \\
\bottomrule
\end{tabular}
\label{tab:fleurs}
\end{table}

\subsection{Main Results}
We start with evaluating \ourmethod{} on the VoxPopuli dataset. The results are presented in Table \ref{tab:voxpopuli}. \ourmethod{} outperforms the Whisper baselines by a notable margin, over $6\%$ w.r.t the best performing baseline. Moreover, \ourmethod{} outperforms the baselines in $15$ of $16$ languages with $\sim 10\times$ fewer parameters.
\label{exp:voxpopuli}
We follow the protocol from~\cite{shin2022learning, nishu2023flexible, nishu2023matching} and evaluate \ourmethod{} on the LibriPhrase dataset. We use \ourmethod{} pre-trained on VoxPopuli and fine-tune it using $30K$ samples from the  LibriPhrase training set for an additional $10$ epochs. The results are presented in Table ~\ref{tab:libriphrase}.
\ourmethod{} significantly outperforms the current SOTA method on the challenging LibriPhrase Hard benchmark.

\subsection{Generalization to Novel Languages and Datasets}

We further investigate the generalization ability of \ourmethod{} in two ways: Adapting to novel, low-resource languages, and generalizing to datasets that are not utilized during training. For both experiments, we use \ourmethod{} trained on the VoxPopuli dataset from Section~\ref{exp:voxpopuli} \emph{without fine-tuning}. 
First, we evaluate the pre-trained \ourmethod{} on a subset of $4$ languages randomly drawn from Fleurs~\cite{Conneau2022FLEURSFL}, which share the same character set observed during training.
Second, we evaluate how well \ourmethod{} generalizes to unseen utterances from the multilingual LibriSpeech (MLS)~\cite{Panayotov2015LibrispeechAA} dataset. We also report the inference time for each model estimated using a single V$100$ Nvidia GPU, averaged using $1000$ random samples.
The results are presented in Tables~\ref{tab:fleurs} and~\ref{tab:mls}. For MLS, our \ourmethod{}-Small model achieves on-par performance with $\sim 160$x faster inference time, w.r.t the Whisper large-v2 baseline.

\subsection{Ablation Study}
We empirically investigate the effects of distinct negative sampling methods on the performance of our model on the Voxpopuli dataset. The results presented in Table~\ref{tab:neg_sampling} showcase the importance of the proposed negative sampling approach for \ourmethod{} performance.


\begin{table}[H]
\centering
    \caption{\textit{Negative sampling:} Results for the \ourmethod{}-tiny model, trained using different negative sampling approaches, on the VoxPopuli test dataset.}
    \vskip 0.11in
\scriptsize
\begin{tabular}{lcccc}
\toprule
 &  F1 $\uparrow$ & AUC $\uparrow$ & EER $\downarrow$\\
\midrule
Random & $79.37$ & $88.09$ & $19.37$  \\
Random + NK & $81.24$ & $88.79$ & $18.80$\\
Random + NK + Cat. & $84.99$ & $91.20$ & $15.91$\\
Random + NK + Cat. + Swap & $\mathbf{92.87}$ & $\mathbf{97.78}$ & $\mathbf{7.16}$ \\
\bottomrule
\end{tabular}
\label{tab:neg_sampling}
\end{table}

\section{Conclusion}
In this paper, we present \ourmethod{}, a novel approach for keyword spotting. 
Our model employs a keyword adaptive instance normalization layer which allows the model to adapt to novel keywords during inference.
Along with the introduced negative keyword mining approach, \ourmethod{} achieves state-of-the-art results in challenging open-vocabulary multilingual setups.
Furthermore, we demonstrate the generalization abilities of \ourmethod{} to adapt to unseen languages and datasets. The impressive achievements of \ourmethod{} not only advance the current state-of-the-art but also pave the way for future research in this domain.

\bibliographystyle{IEEEbib}
\bibliography{refs}

\end{document}